\documentclass[traditabstract]{aa}

\usepackage{times}
\usepackage{epsfig,graphicx,pdflscape}
\usepackage{float,longtable,comment}
\usepackage{rotating,amsmath}
\usepackage{amssymb}
\usepackage{natbib}
\newcommand{\kms}{{{km\,s}$^{-1}$\,}}

\newcommand{\vsini}{$v_\mathrm{e} \sin i$}
\newcommand{\Msun}{\,$\mathrm{M}_\odot$}
\newcommand{\pp}{$\phantom{1}$}
\newcommand{\mm}{$\phantom{-}$}
\defcitealias{eva11}{Paper~I}
\defcitealias{san13}{Paper~VIII}

\begin{document}

\title{The VLT-FLAMES Tarantula Survey}
\subtitle{XXII. Multiplicity properties of the B-type stars}  

\author{P.~R.~Dunstall\inst{1}, P.~L.~Dufton\inst{1}, H.~Sana\inst{2}, C.~J.~Evans\inst{3}, 
I.~D.~Howarth\inst{4}, S. Sim\'on-D\'{i}az\inst{5,6} \\
S.~E.~de~Mink\inst{7}, N.~Langer\inst{8}, 
J. Ma\'{i}z Apell\'{a}niz\inst{9}, W.~D.~Taylor\inst{3}}

\institute{Astrophysics Research Centre, School of Mathematics \& Physics, Queen's University Belfast, BT7 1NN, Northern Ireland, UK	
\and ESA\,/\,Space Telescope Science Institute, 3700 San Martin Drive, Baltimore, MD 21218, USA	
\and UK Astronomy Technology Centre, Royal Observatory Edinburgh, Blackford Hill, Edinburgh, EH9 3HJ, UK        
\and Dept. of Physics \& Astronomy, University College London, Gower Street, London, WC1E 6BT, UK
\and Instituto de Astrof\'{i}sica de Canarias, E-38205 La Laguna, Tenerife, Spain        
\and Departamento de de Astrof\'{i}sica, Universidad de La Laguna, E-38205 La Laguna, Tenerife, Spain
\and Astronomical Institute Anton Pannekoek, Amsterdam University, Science Park 904, 1098 XH, Amsterdam, The Netherlands
\and Argelander-Institut f\"{u}r Astronomie der Universit\"{a}t Bonn, Auf dem H\"{u}gel 71, 53121 Bonn, Germany
\and Departamento de Astrof\'{i}sica, Centro de Astrobiolog\'{i}a (INTA-CSIC), Campus ESA, Apartado Postal 78, 28\,691 Villanueva de la Ca\~{n}ada, Madrid, Spain}

\offprints{P.~L.~Dufton at P.Dufton@qub.ac.uk}

\date{Accepted: 21 May 2015 }

\abstract{We investigate the multiplicity properties of 408 B-type
  stars observed in the 30~Doradus region of the Large Magellanic
  Cloud with multi-epoch spectroscopy from the VLT-FLAMES Tarantula
  Survey (VFTS). We use a cross-correlation method to estimate
  relative radial velocities from the helium and metal absorption
  lines for each of our targets. Objects with significant
  radial-velocity variations (and with an amplitude larger than
  16~km\,s$^{-1}$) are classified as spectroscopic binaries. We find
an observed spectroscopic binary fraction (defined by periods of
  $<$10$^{3.5}$\,d and mass ratios $>$\,0.1) for the B-type stars,
$f_{\rm B}$(obs)\,$=$\,0.25\,$\pm$\,0.02, which appears constant
across the field of view, except for the two older clusters
(Hodge\,301 and SL\,639). These two clusters have significantly
lower binary fractions of 0.08\,$\pm$\,0.08 and 0.10\,$\pm$\,0.09,
respectively.  Using synthetic populations and a model of our observed
epochs and their potential biases, we constrain the intrinsic
multiplicity properties of the dwarf and giant (i.e. relatively
unevolved) B-type stars in 30~Dor.  We obtain a present-day binary
fraction $f_{\rm B}$(true)\,$=$\,0.58\,$\pm$\,0.11, with a flat period
distribution. Within the uncertainties, the multiplicity properties of
the B-type stars agree with those for the O stars in 30~Dor from the
VFTS.}


\authorrunning{P.~R.~Dunstall et al.}
\titlerunning{Multiplicity of B-type stars in 30 Dor.}

\maketitle

\section{Introduction}\label{s_intro}

The presence of a close companion can significantly influence the
evolution of massive stars \citep[e.g.][]{pod92, lan08, eld11}, but
the prevalence of mass transfer in such systems has only recently been
realised.  A consequence of the high spectroscopic binary fractions
inferred for massive stars in the Galaxy
\citep[e.g.][]{mas09,san08,san09,san11,kk12,chini12,sota14} is that
the majority of massive stars may be born as part of, and reside
within, multiple systems \citep{kf07,san12}. Additionally, stars that
currently appear to be single may have evolved from binary systems
\citep{sdm14}, with important consequences on their inferred
evolutionary histories \citep{dem11,dem13}.

The multiplicity fraction of B-type stars has been investigated in
some Galactic clusters \citep[e.g.][]{rab96} but, to date, most
studies of the multiplicity of high-mass stars have focused on the
more massive O-type objects \citep[e.g.][and references
therein]{se11}. Substantial spectroscopic samples of B-type stars
in the Galaxy and Magellanic Clouds are available from large
spectroscopic programmes \citep[e.g.][]{eva05, eva06, mar06, mart07},
but they have lacked sufficient time sampling to investigate the role of
binarity in the observed populations.

One of the primary motivations for the VLT-FLAMES Tarantula Survey
\citep[VFTS,][hereafter \citetalias{eva11}]{eva11} is to investigate
the multiplicity properties of the massive stars in the 30~Doradus
region of the Large Magellanic Cloud (LMC). To this end, the VFTS
obtained multi-epoch spectroscopy of $\sim$800 O- and early B-type
stars. The observed spectroscopic binary fraction, $f_{\rm O}$(obs),
of the 360 O-type stars in the VFTS was found to be $f_{\rm
  O}$(obs)\,$=$\,35\,$\pm$\,3\% \citep[][hereafter
\citetalias{san13}]{san13}; once the observational biases were taken
into account, this gave an intrinsic binary fraction, $f_{\rm
  O}$(true)\,$=$\,51\,$\pm$\,4\%.

Here we investigate the multiplicity properties of the B-type stars
observed by the VFTS. One of our objectives is an estimate of their
present-day observed spectroscopic binary fraction, $f_{\rm B}$(obs),
where $f$ defines the fraction of targets where Doppler shifts are
detected (rather than the total fraction of all stars), probing
separations with orbital periods of up to approximately 10 years
\citep[e.g.  see Fig.~1 from][]{se11}. For consistency with the
analysis by \citet{san12} and in \citetalias{san13}, we adopt the same
definition of a spectroscopic binary as a system with a period of
$<$10$^{3.5}$\,d and a mass ratio of $>$\,0.1. The present-day binary
fraction can then be used to constrain the minimum binary fraction of
B-type stars at their birth (with initial masses of $\gtrsim$
8\,\Msun).  Ultimately, in combination with the results for the O-type
stars (from \citetalias{san13}), our aim is to determine the fraction
of massive stars that might be affected by binary interaction during
their lives, which will influence their evolution and the properties
of their final explosions. Indeed, assuming the form of the stellar
mass-function from \citet{kro01}, around two thirds of the progenitors
of core-collapse supernovae were probably early B-type stars when on
the main sequence.

After a brief summary of the observations in Section~\ref{s_obs}, we
describe the methods used to identify spectroscopic binaries in
Section~\ref{s_analysis}. We present our results in
Section~\ref{s_binary} and investigate the intrinsic properties of the
binary population in Section~\ref{s_model}. Concluding remarks are
given in Section~\ref{s_conclusions}.

\section{Observations}\label{s_obs}

The VFTS spectra were obtained with the Fibre Large Array
Multi-Element Spectrograph \citep[FLAMES,][]{pas02} on the Very Large
Telescope (VLT). Classifications for the 438 B-type stars observed in
the VFTS were given by \citet{eva15}. These are located in the main
clusters in the 30~Dor region (i.e. NGC\,2070, NGC\,2060, Hodge\,301,
SL\,639) and the local field population (see Fig.~4 from
\citeauthor{eva15}). All of the spectra were obtained using the
fibre-fed Medusa--Giraffe mode of FLAMES, so the sample does not
include stars in R136, the young massive cluster at the core of
30~Dor, which is too densely populated for effective use of the Medusa
fibres.

This paper presents a radial-velocity (RV) analysis of the multiple
observations of the B-type stars with the LR02 setting of the Giraffe
spectrograph\footnote{The LR02 observations provided coverage of 3960 to 4560\,\AA\ at a
spectral resolving power of $\sim$6\,500. Observations were
  also obtained using the LR03 and HR15N settings, but with
  considerably limited time sampling (cf. the LR02 data) so these were
  not considered further here.}.
The observations spanned 10--12 months for most targets, with an
extended baseline of 22 months for 31 targets, due to a reobservation
for operational reasons. Two or three consecutive exposures were
obtained on any given night, with the longer-term sampling designed to
optimise detection of binaries with periods up to $\sim$200\,d, given
scheduling constraints. \citepalias[The detection probability
decreases significantly for longer-period systems; see e.g.\ Fig.~8
of][]{san13}. Further details regarding target selection, data
reduction, and the observation dates are given in \citetalias{eva11}.

Absolute RVs for the B-type sample were provided by \citet{eva15}, 
which are either the mean of estimates from each observation or, where
binary motion was suspected, the estimate from a particular observation. 
For 30 stars the spectra were not of sufficient quality to allow
reliable estimates of the absolute RVs, and these stars have been excluded
from our sample. As these span a range of spectral types and
morphologies (including examples of Be-type stars), they should not
unduly influence the results. A further five targets are the
double-lined binaries identified by \citet{eva15}, namely: VFTS\,112,
199, 240, 637, and 698\footnote{A detailed discussion of VFTS\,698 was
  given by \citet{dun12}. The SB2 system VFTS\,652 was
  omitted from the \citeauthor{eva15} sample, and hence from this paper,
  because of the O-type classification of the secondary \citep{wal14}.
  The ten B0-type stars classified by \citet{wal14} were also
  discussed by \citet{eva15}, so are included in this paper.}.  It was
challenging to obtain multi-epoch RV estimates to the same standard as
those used for the analysis in Section~\ref{s_analysis}, but their
binary natures were easily revealed by variations in their spectra
(indicative of binary motion, rather than simply composite spectra).

Therefore, in total we investigated the multiplicity of 408 stars,
comprised of 361 dwarfs and giants (i.e. relatively unevolved
objects) and 47 supergiants \citep[defined as having a logarithmic
gravity, $\log$\,g\,$\le$\,3.3\,dex, see][]{mce15}; hereafter these are
referred to as the `unevolved' and `supergiant' samples.

\section{Radial velocities}\label{s_analysis}

\subsection{Methodology}\label{s_rvmeasure}

We began by examining the 403 single-lined targets for evidence of RV
variability.  An important aspect of RV studies in early-type stars
concerns the choice of diagnostic lines.  Our preferred features for
B-type stars would be their intrinsically narrow metal absorption
lines. However, the large variation in the intensity of these lines
with spectral type means they would not provide a homogeneous set of
estimates on their own. We therefore employed both helium and metal
absorption lines (see Table~\ref{t_lines}) to obtain a range of
independent RV estimates, as permitted by the signal-to-noise (S/N)
ratio of the data and the spectral type of each target.

For each of the lines available from Table~\ref{t_lines}, relative RV
estimates ($v_{i}$) were obtained for each LR02 spectrum ($i$) of a
given target from cross-correlation of the data with an adopted
reference spectrum (typically that for the object with the largest
number of counts in the continuum around 4200 \AA). This approach has
been widely used \citep[see e.g.][and references therein]{mar15,
  man15, rob14}, with our methodology being similar to that used by
\citet{mar15}.

The spectral region used in the cross-correlation analysis depends on
the width of the spectral feature being considered.  In turn this
depends on both its intrinsic width (e.g. diffuse helium lines will be
far broader than non-diffuse helium lines or metal lines) and on the
stellar projected rotational velocity. The projected rotational
velocity estimates from \citet{duf13} for the subset of our sample
which have no significant radial-velocity shifts cover a wide range.
The distribution is bi-modal with peaks at \vsini $\simeq$ 20 and 180
\kms, coupled with a long tail to around 400 \kms. The latter would
equate to a full-width-half maximum rotational broadening of about
6\AA\ (depending on the stellar wavelength). Given this wide variation
in the width of spectral features, the choice of spectral regions for
each line was made subjectively. The principle criteria were to
minimise the region selected (to maximise the cross-correlation
signal), whilst ensuring both complete coverage of the feature and
sufficient continuum to fully characterise the cross-correlation
function.

\begin{figure*}
\begin{center}
\includegraphics[width=15.75cm]{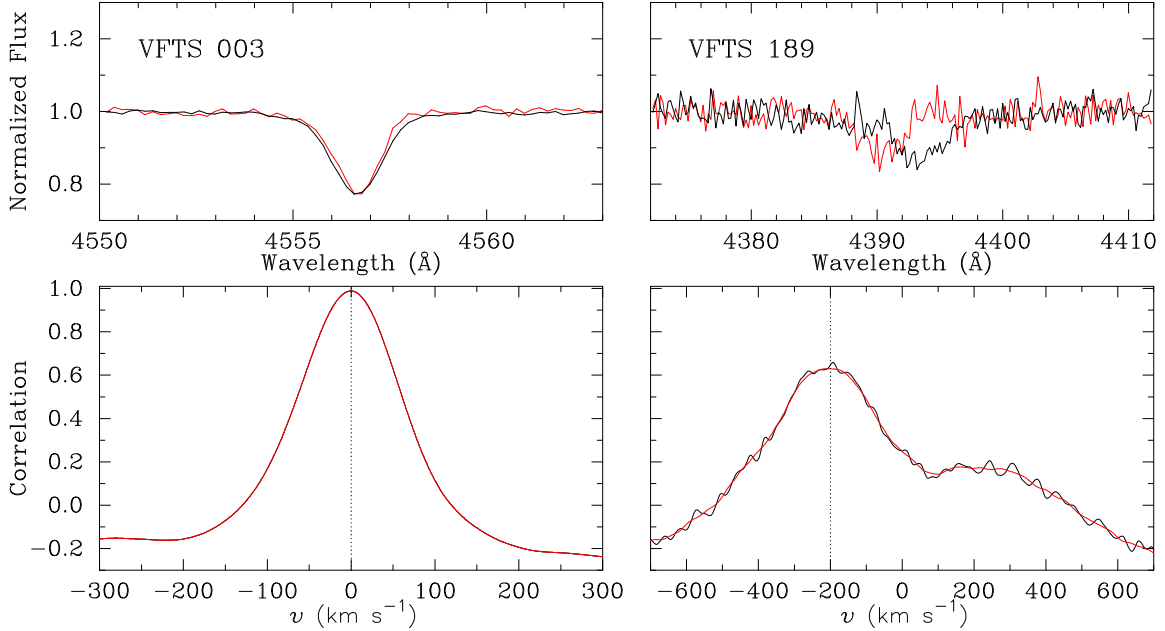} 
\caption{Examples of the methodology to estimate radial-velocity
  shifts between the spectra.  {\it Left-hand panels:} VFTS\,003 is a
  relatively bright narrow-lined supergiant. The upper panel shows the
  \ion{Si}{iii} 4553\AA\ line in the template spectrum
  (MJD\,$=$\,54804.148, black line) and at MJD\,$=$\,55108.309 (red).
  The lower panel shows the original (black line) and smoothed (red)
  cross-correlation function in velocity space; no significant
  velocity shifts were found for this target. {\it Right-hand panels:}
  VFTS\,189 is a relatively faint, broad-lined, main-sequence star.
  The upper panel shows the \ion{He}{i} 4388\AA\ line in the template
  spectrum (MJD\,$=$\,54867.160, black) and at MJD\,$=$\,54824.242
  (red). The lower panel shows the original and smoothed
  cross-correlation functions in velocity space; the estimated radial-velocity
  shift between these was $\sim$200\,\kms.}
\label{vr_examples}
\end{center}
\end{figure*}

The position of the maximum could be affected by spurious
structure in the cross-correlation function, so our method was
informed by the projected equatorial rotation velocities (\vsini) for
each star from \citet{duf13}. For supergiants and narrow-lined spectra
(\vsini\,$\le$\,150\,\kms) the cross-correlation function was smoothed
by a 5-point moving average (equivalent to $\sim$50\,\kms in velocity
space).  For stars rotating more rapidly, a 55-point moving average
was used, to militate against the increased influence of random noise
in the cross-correlation function; this method was tested for the
narrow-lined spectra giving, in general, identical results.

The procedure is illustrated for two cases in Fig.~\ref{vr_examples},
viz. the \ion{Si}{iii} 4553\AA\ line in the narrow-lined bright
supergiant VFTS\,003 (B1 Ia$^+$) and the \ion{He}{i} 4388\AA\ diffuse
line in the relatively rapidly rotating (\vsini $\simeq$ 212 \kms) and
fainter main-sequence star, VFTS\,189 (B0.7: V); as such they
illustrate the range in quality of the available observational data.

The relatively narrow absorption lines and high S/N of the supergiant
spectra meant that all of the metal and helium lines listed in
Table~\ref{t_lines} could be used for the RV analysis; this was also
possible for $\sim$20\% of the (non-supergiant) narrow-lined objects.
For more-rapidly-rotating stars (\vsini\,$\ge$\,150\,\kms), only the
helium lines were generally available. 

\begin{table}
\caption{Absorption lines used to estimate relative radial velocities for the B-type
spectra.}\label{t_lines}
\begin{center}
\begin{tabular}{lcl}
\hline
Ion & $\lambda$ (\AA) & Type \\
\hline
\ion{O}{ii} & 4080 & Group \\
\ion{He}{i} & 4009 & Singlet \\
\ion{He}{i} & 4026 & Triplet \\
\ion{He}{i} & 4121 & Triplet \\
\ion{He}{i} & 4144 & Singlet \\
\ion{He}{i} & 4169 & Singlet \\
\ion{O}{ii} & 4317 & Doublet \\
\ion{He}{i} & 4388 & Singlet \\
\ion{O}{ii} & 4414 & Doublet \\
\ion{He}{i} & 4438 & Singlet \\
\ion{He}{i} & 4471 & Triplet \\
\ion{Si}{iii} &	4553 & Singlet \\
\hline
\end{tabular}
\tablefoot{The \ion{O}{ii} $\lambda$4080 group comprises a number
  of lines extending over the 4069-4095\,\AA\ region.}
\end{center}
\end{table}

The relative RVs for each observation of 370 of our targets are
presented in Table~\ref{RVs}; the remaining 33 targets were affected
by nebular contamination and are discussed in the next section, while
a RV analysis was not attempted for the five targets classified as SB2
systems. We note that non-zero estimates arise from cross-correlation of
the template spectra with themselves, with a systematic median offset
of $-$0.28\,\kms\ arising from the fits to the cross-correlation
function.  In the context of our analysis of {\em relative} velocities
and the adopted criteria to identify binaries (Section~\ref{s_bc})
such measurement uncertainties are not significant.

\begin{table*} 
  \caption{Relative radial-velocity (RV) estimates for the available lines in each observation (cf. the adopted
template spectrum for each star).  Values for the first three VFTS targets are shown; the full 
table is available online.}\label{RVs}
{\tiny 
\begin{center}
\begin{tabular}{llrrrrrrrrrrrr}
\hline 
VFTS & MJD & \multicolumn{12}{c}{Relative radial velocities [\kms]}\\
\hline
& & \ion{Si}{iii} & \ion{O}{ii} & \ion{O}{ii} & \ion{O}{ii} & \ion{He}{i} & \ion{He}{i} 
& \ion{He}{i} &  \ion{He}{i} & \ion{He}{i} & \ion{He}{i} & \ion{He}{i} & \ion{He}{i}\\
$\phantom{Star}$& & $\lambda$4553 & $\lambda$4080 & $\lambda$4317 & $\lambda$4414 & $\lambda$4026 & 
$\lambda$4471 & $\lambda$4009 & $\lambda$4144 & $\lambda$4388 & $\lambda$4121 & $\lambda$4438 & $\lambda$4169 \\
\hline
001 & 54815.254 & ... & ... & ... & ... & $-$25.48 & ... & ... & ... & 10.55 & ... & ... & ... \\
001 & 54815.277 & ... & ... & ... & ... & $-$6.33 & ... & ... & ... & $-$31.42 & ... & ... & ... \\
001 & 54815.305 & ... & ... & ... & ... & $-$13.05 & ... & ... & ... & $-$1.63 & ... & ... & ... \\
001 & 54815.328 & ... & ... & ... & ... & $-$22.53 & ... & ... & ... & $-$28.72 & ... & ... & ... \\
001 & 54859.211 & ... & ... & ... & ... & $-$9.27 & ... & ... & ... & $-$9.76 & ... & ... & ... \\
001 & 54859.230 & ... & ... & ... & ... & 6.93 & ... & ... & ... & $-$19.24 & ... & ... & ... \\
001 & 54891.062 & ... & ... & ... & ... & $-$32.84 & ... & ... & ... & $-$13.82 & ... & ... & ...  \\
001$^\ast$ (S) & 54891.082 & ... & ... & ... & ... & $-$0.44 & ... & ... & ... & $-$0.28 & ... & ... & ... \\
\hline
003 & 54804.082 & 0.76 & 2.83 & $-$0.57 & 0.92 & $-$0.44 & 1.05 & $-$0.05 & $-$1.98 & 2.43 & $-$0.56 & $-$1.86 & ...\\
003 & 54804.105 & 0.76 & 1.33 & 0.79 & 0.92 & $-$1.91 & 1.05 & 1.44 & $-$0.55 & 1.08 & 0.88 & 0.81 & ...\\
003 & 54804.125 & 0.76 & 2.83 & 0.79 & 0.92 & 1.04 & 1.05 & 1.44 & $-$0.55 & 1.08 & 0.88 & 0.81 & ...\\
003$^\ast$ (V) & 54804.148 & $-$0.54 & $-$0.17 & $-$0.57 & $-$0.42 & $-$0.44 & $-$0.28 & $-$0.05 & $-$0.55 & $-$0.28 & $-$0.56 & $-$0.53 & ...\\
003 & 54804.168 & 0.76 & 1.33 & 0.79 & $-$0.42 & $-$0.44 & 1.05 & $-$0.05 & $-$1.98 & $-$0.28 & $-$0.56 & $-$0.53 & ...\\
003 & 54804.191 & 0.76 & $-$0.17 & 0.79 & $-$0.42 & 1.04 & 1.05 & 1.44 & $-$0.55 & $-$0.28 & $-$0.56 & $-$0.53 & ...\\
003 & 54836.219 & 4.65 & 2.83 & 2.16 & 0.92 & 8.40 & 10.35 & 2.92 & 2.30 & 3.78 & 2.31 & 6.14 & ...\\
003 & 54836.238 & 4.65 & 2.83 & 2.16 & 0.92 & 8.40 & 10.35 & 4.41 & 2.30 & 5.14 & 2.31 & 0.81 & ...\\
003 & 54836.266 & 3.35 & 2.83 & 2.16 & 0.92 & 8.40 & 10.35 & 4.41 & 2.30 & 5.14 & 2.31 & 0.81 & ...\\
003 & 54836.285 & 4.65 & 2.83 & 2.16 & 0.92 & 6.93 & 11.67 & 2.92 & 2.30 & 3.78 & 2.31 & 3.47 & ...\\
003 & 54867.086 & 13.73 & 8.84 & 11.72 & 7.64 & 17.24 & 19.64 & 13.33 & 12.28 & 14.62 & 10.92 & 14.14 & ...\\
003 & 54867.109 & 15.02 & 10.34 & 13.09 & 6.29 & 17.24 & 20.97 & 13.33 & 12.28 & 13.26 & 8.05 & 7.47 & ...\\
003 & 55108.309 & $-$0.54 & $-$0.17 & $-$0.57 & $-$0.42 & $-$0.44 & $-$1.61 & 1.44 & $-$1.98 & $-$2.99 & 0.88 & $-$3.19 & ...\\
\hline
005 & 54815.254 & ... & ... & ... & ... & $-$7.80 & ... & ... & 6.58 & $-$1.63 & ... & ... & ... \\
005 & 54815.277 & ... & ... & ... & ... & $-$0.44 & ... & ... & $-$0.55 & 2.43 & ... & ... & ... \\
005 & 54815.305 & ... & ... & ... & ... & 18.71 & ... & ... & $-$17.67 & 3.78 & ... & ... & ... \\
005 & 54815.328 & ... & ... & ... & ... & $-$7.80 & ... & ... & 16.56 & $-$1.63 & ... & ... & ... \\
005 & 54859.211 & ... & ... & ... & ... & $-$1.91 & ... & ... & 17.99 & 24.10 & ... & ... & ... \\
005 & 54859.230 & ... & ... & ... & ... & $-$10.75 & ... & ... & 8.00 & 13.26 & ... & ... & ... \\
005 & 54891.062 & ... & ... & ... & ... & $-$12.22 & ... & ... & $-$6.26 & $-$1.63 & ... & ... & ... \\
005$^\ast$ (S) & 54891.082 & ... & ... & ... & ... & $-$0.44 & ... & ... & $-$0.55 & $-$0.28 & ... & ... & ... \\
005 & 55113.309 & ... & ... & ... & ... & $-$13.69 & ... & ... & 0.87 & 21.39 & ... & ... & ... \\
005 & 55113.328 & ... & ... & ... & ... & $-$0.44 & ... & ... & 10.86 & 11.91 & ... & ... & ... \\
\hline
\end{tabular}
\tablefoot{Celestial coordinates and optical photometry for the VFTS
  stars are given in Table~5 of Paper~I. Observations indicated
  by the $^\ast$ qualifier to the first column were the template
  spectra for the RV estimates. The subsequent letters in parentheses
  indicate the binary status from this work: (B)\,$=$\,binary;
  (S)\,$=$\,single; (V)\,$=$\,RV variable; see Section~\ref{s_bc} for
  further details.}
\end{center}
}
\end{table*}

\subsection{Nebular contamination}\label{s_nebular}
There is considerable nebulosity across the 30~Dor region, which
contaminates the FLAMES spectroscopy to varying degrees.  This can
complicate the RV estimates by the introduction of additional
structure (centred at $\Delta$RV\,$\simeq$\,0\,\kms) in the
cross-correlation profiles of the helium lines.  For the majority of
our targets the nebular emission was sufficiently weak that any
associated structure in the cross-correlation function was minor, and
robust RV estimates were possible.  However, the contamination was
more significant for 33 of our targets, so we employed a modified
method where all exposures for a given date were merged to improve the
S/N, and then analysed as for the other spectra. The relative
velocities from this approach for these targets are given in
Table~\ref{RVs_single}.

\subsection{Criteria for binarity}\label{s_bc}
Our strategy for binary detection relies on statistical criteria, with
the significance threshold adopted such that it limited the number of
false positives. Thus, while some binaries may remain undetected, this
systematic approach enables modelling of the observational biases.  For
consistency, we adopted similar criteria for binarity to those used in
the analysis of the O-type stars \citepalias{san13}, such that an object
is considered a spectroscopic binary if at least one pair of RV
estimates satisfies simultaneously
\begin{equation}
\frac{|v_{i} - v_{j}|}{\sqrt{\sigma_{i}^2 + \sigma_{j}^2}} > 
4.0~~~{\rm and}~~~|v_{i} - v_{j}| > \Delta{\rm RV}_{\rm min}, \label{criteria}
\end{equation}
where $v_{i}$ is the mean relative RV for exposure $i$ with respect to
the template, and $\sigma_{i}$ is its associated standard deviation.
The choice of 4.0 for the confidence threshold for a detection was
guided by the number of false positives one may statistically expect
given the sample size, the uncertainties of the RV estimates, and the
number of RV pairs.  Using Monte Carlo simulations, adopting 4.0 leads
to fewer than 0.2 false detections for the whole sample, which is
sufficient for our purposes. Furthermore, this threshold value is
identical to that adopted in \citetalias{san13}, ensuring a consistent
approach.

\begin{figure}
\begin{center}
\includegraphics[width=7.5cm]{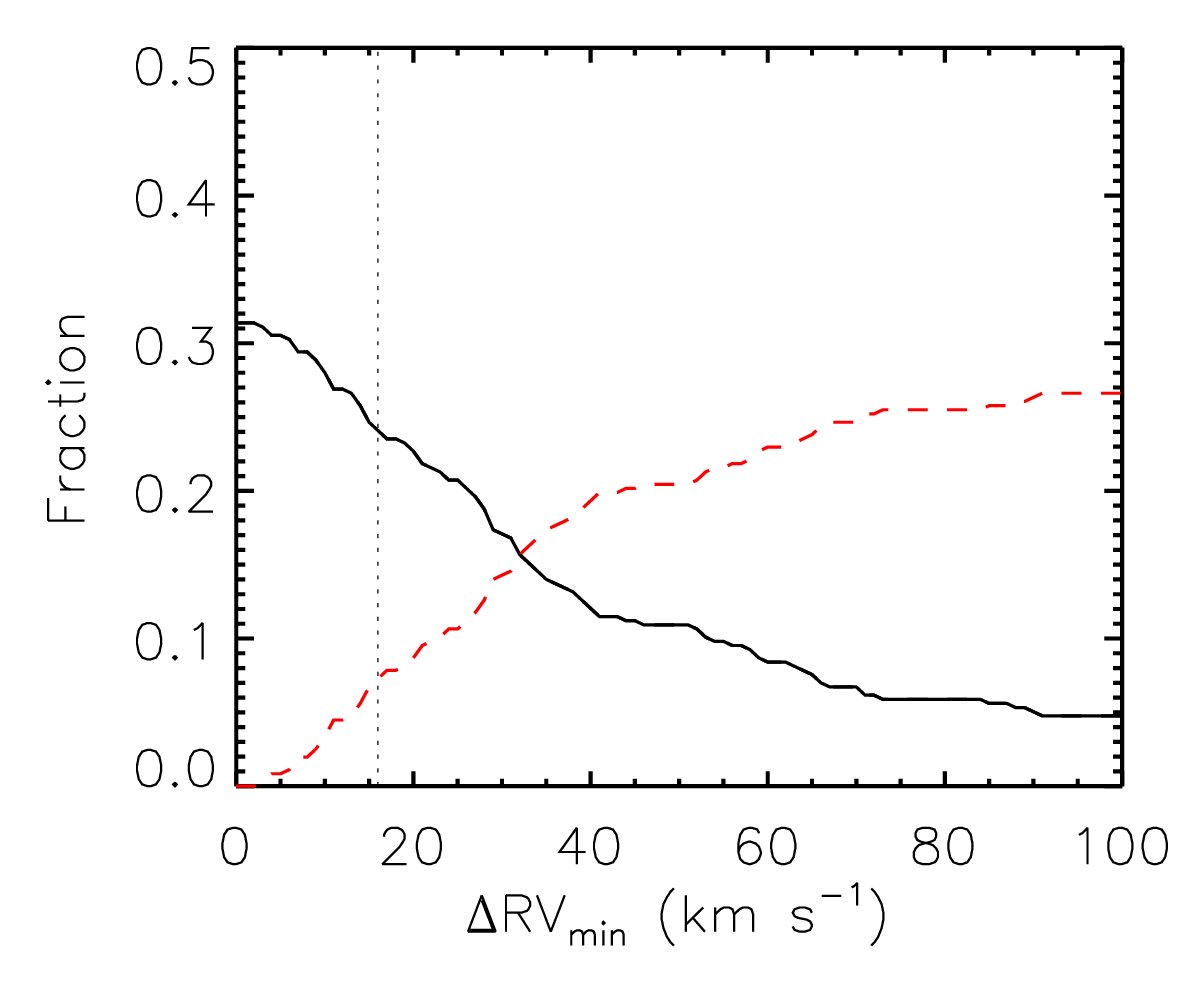}
\includegraphics[width=7.5cm]{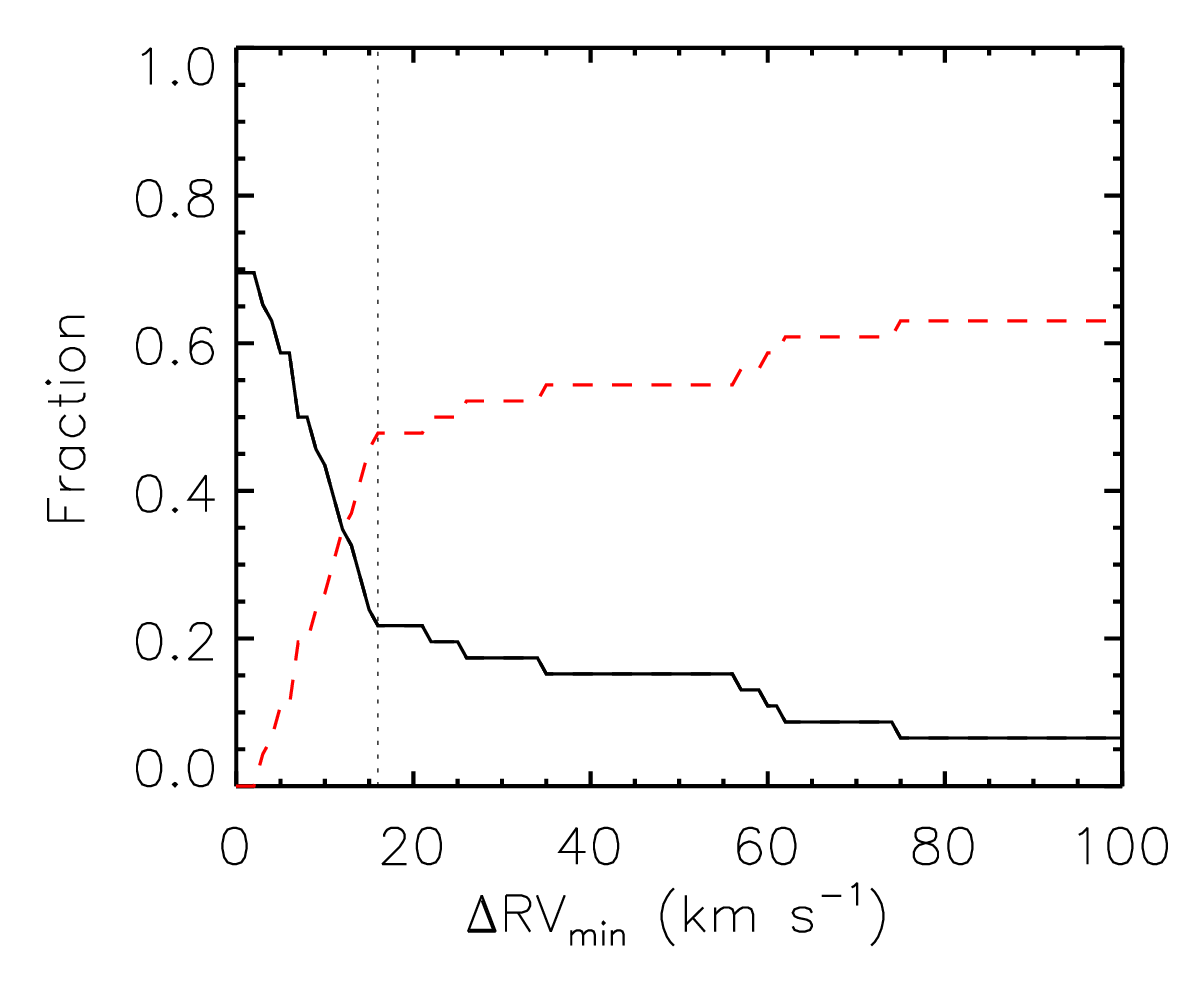}
\caption{{\it Upper panel:} Inferred binary fraction as a function of
  the adopted threshold velocity ($\Delta$RV$_{\rm min}$) for the
  unevolved (dwarf/giant) B-type stars in our sample (solid line).
  The dashed line shows the corresponding fraction that would have
  been classified as binary candidates but failed the $\Delta$RV$_{\rm
    min}$ criterion. {\it Lower panel:} As for the upper panel, but
  for the supergiants. In both plots the vertical dotted line
  indicates the adopted threshold of 16\,\kms.}\label{f_dRV}
\end{center}
\end{figure}

The choice of $\Delta$RV$_{\rm min}$ was particularly important for
the supergiants, for which precise RVs could be determined given their
high S/N and relatively narrow-lined profiles.  Small ranges of
variations ($\simeq$\,5-10\,\kms) could indicate a binary companion,
but may also result from atmospheric activity or pulsations
\citep[e.g.][]{sher25}.

The percentage of binary candidates identified for different values of
$\Delta$RV$_{\rm min}$ is illustrated in Fig.~\ref{f_dRV}. For the
unevolved stars, no clear break is seen.  However, the fraction of
RV-variable objects for the supergiants presents a clear kink around
$\Delta$RV$_{\rm min}$\,$=$\,16\,\kms\, with samples plausibly
dominated by binarity and intrinsic variability above and below the
kink, respectively.  This is a slightly lower threshold than that
adopted in \citetalias{san13} for the O-type stars (i.e.
$\Delta$RV$_{\rm min}$\,$=$\,20\,\kms), but is consistent with the
lower masses of early B-type stars (cf.  those for O stars\footnote{For a 
given mass ratio and orbital period, the semi-amplitude of the RV signal
for the primary is $\propto$\,$M^{1/3}$.}).

Objects fulfilling both criteria of Eq.~\ref{criteria} (with
$\Delta$RV$_{\rm min}$\,$=$\,16\,\kms) are considered as spectroscopic
binaries.  Those fulfilling the first condition of Eq.~\ref{criteria},
but with a maximum $\Delta$RV in the range of 5 to 16\,\kms\ are
identified as `RV variables', with the variations attributable to
either binarity or intrinsic variability.  The remaining objects are
presumed to be single stars for the purposes of the following
discussion.\footnote{As discussed in Section~\ref{s_obs}, 30 targets
  were excluded from our sample because absolute RV values could not
  be measured.  However, relative velocity estimates (often only from
  one feature) were determimed for 15 of these targets, with tentative
  evidence of RV variations found in seven stars: VFTS\,155, 377, 391
  434, 442, 742, and 821 \citep[and flagged as such by][]{eva15}, with
  no significant variations seen in the remaining eight stars:
  VFTS\,366, 407, 408, 462, 573, 653, 689, and 854. Given the
  inhomogeneity of these estimates in terms of the lines and
  observations these stars are not considered further, but we note
  these results for completeness (cf. \citeauthor{eva15}).}

\begin{table} 
  \caption{Relative radial-velocities (RV) for spectra with
    significant nebular contamination but which estimates were possible from combining the spectra
    from each night (see Section~\ref{s_nebular}). Values for the first three entries are shown;
    the full table is available online.}\label{RVs_single}
\begin{center}
\begin{tabular}{llrrrr}
\hline 
VFTS & MJD & \multicolumn{4}{c}{Relative radial velocities [\kms]}\\
\hline
& &  \ion{He}{i} & \ion{He}{i} & \ion{He}{i} &  \ion{He}{i} \\
& &  $\lambda$4388 & $\lambda$4144 & $\lambda$4026 & $\lambda$4471 \\
\hline
004$^\ast$ (S) & 54767 &\mm\pp0.39 &\mm\pp0.44 &\mm\pp0.29 &\pp$-$1.42 \\
004  & 54827 &  $-$14.28 &  $-$15.10 &  $-$12.47 &  $-$71.61 \\
004  & 54828 &  $-$39.97 &  $-$59.7 &  $-$23.65 &  $-$14.02 \\
004  & 54860 &  $-$10.62 &\pp$-$3.45 &  \mm17.85 &\mm\pp0.37 \\
004  & 54886 &  $-$60.14 &  $-$40.36 &  $-$12.47 &  $-$23.02 \\
004  & 55114 &  \mm53.58 &\pp$-$3.45 &  $-$58.75 &  $-$28.42 \\
\hline
010  & 54767 &  \mm26.80 &\pp\mm5.19 &  $-$14.43 &\pp\mm6.94 \\
010  & 54827 &\mm\pp9.20 &  $-$12.24 &  $-$26.21 &  $-$32.39 \\
010$^\ast$ (S) & 54860 &\pp$-$0.28 &\pp\mm0.44 &\pp$-$1.17 &\pp$-$0.34 \\
010  & 54886 &  $-$39.55 &  \mm40.07 &\pp$-$8.54 &     \ldots \\ 
010  & 55114 &\mm\pp4.88 &  \mm41.65 &  $-$17.38 &     \ldots \\
\hline
054$^\ast$ (S) & 54767 &\pp$-$0.28 &\pp$-$0.27 &\pp\mm6.93 &\pp$-$0.28 \\
054  & 54827 &  \mm20.03 &  $-$28.95 &  $-$15.16 &  $-$18.88 \\
054  & 54860 &  \mm49.82 &  \mm24.11 &\pp\mm8.40 &  \mm19.64 \\
054  & 54886 &  $-$21.94 &  $-$23.22 &  $-$60.82 &  $-$37.47 \\
054  & 55114 &  $-$21.94 &  \mm31.28 &  $-$65.24 &  $-$33.49 \\
\hline
\end{tabular}
\tablefoot{Celestial coordinates and optical photometry for the VFTS
  stars are given in Table~5 of Paper~I. Observations indicated by the
  $^\ast$ qualifier to the first column were the template spectra for
  the RV estimates. The subsequent letters in parentheses indicate the
  binary status from this work: (B)\,$=$\,binary; (S)\,$=$\,single;
  (V)\,$=$\,RV variable; see Section~\ref{s_bc} for further details.}
\end{center}
\end{table}

\section{The binary sample}\label{s_binary}

\subsection{Observed binary fraction}\label{s_bf}

Employing the criteria in Eq.~\ref{criteria} to analyse the RV
estimates in Tables~\ref{RVs} and \ref{RVs_single}, combined with the
SB2 systems, we found 90 binaries in the unevolved sample and 11
binaries in the supergiants. The observed binary fractions are
therefore $25 \pm 2$\%\ and $23 \pm 6$\%, respectively, where binomial
statistics have been used to compute the error bars
\citep[see][]{san09}. An additional 23 unevolved stars and 17
supergiants were detected as RV variables, comprising $6\pm1$\%\ and
$36\pm7$\%\ of their respective samples. The lower limit for the
spectroscopic binary fraction for the entire VFTS B-type sample ({\em
  including} the supergiants and SB2 systems) is $f_{\rm
  B}$(obs)\,$=$\,25\,$\pm$\,2\%, with RV variables accounting for a
further $10\pm1$\%\ of the stars.

\subsection{Supergiants}\label{s_supergiants}

The effects of macroturbulence in blue supergiants appear to be
linked to line-profile variations \citep{sim10}, so adoption of
$\Delta$RV$_{\rm min}$\,$=$\,16\,\kms\ in Section~\ref{s_bc} was
particularly important for these objects in the VFTS sample (for which
the uncertainties on the RV estimates are relatively small).  To investigate
the nature of RV variations in the supergiants further we employed the same
procedures as those used by \citet{sim10} to quantify the potential role
of macroturbulence in our targets.

In general, this additional analysis confirmed the classifications
regarding multiplicity using the criteria from Section~\ref{s_bc}.
However, there are four objects (VFTS 420, 423, 541, and 829)
classified as RV variables (with 12\,$<$\,$\Delta$RV\,$<$\,15\,\kms)
for which the results are consistent with minimal macroturbulence,
thereby strengthening the case for orbital RV variations.
Conversely, VFTS\,591, classified as a binary using the above
criteria, has a relatively large contribution from macroturbulence and
thus might be a single star. Revising these classifications would
increase the binary fraction for the supergiants from 23\,$\pm$\,6\%\
to 30\,$\pm$\,7\%, i.e. within the statistical uncertainties,
indicating that uncertainties due to macroturbulence do not appear to
strongly influence the result.

\subsection{Timescales for detected variations}\label{sect: dhjd}

\begin{figure}
\begin{center}
\includegraphics[width=7.5cm]{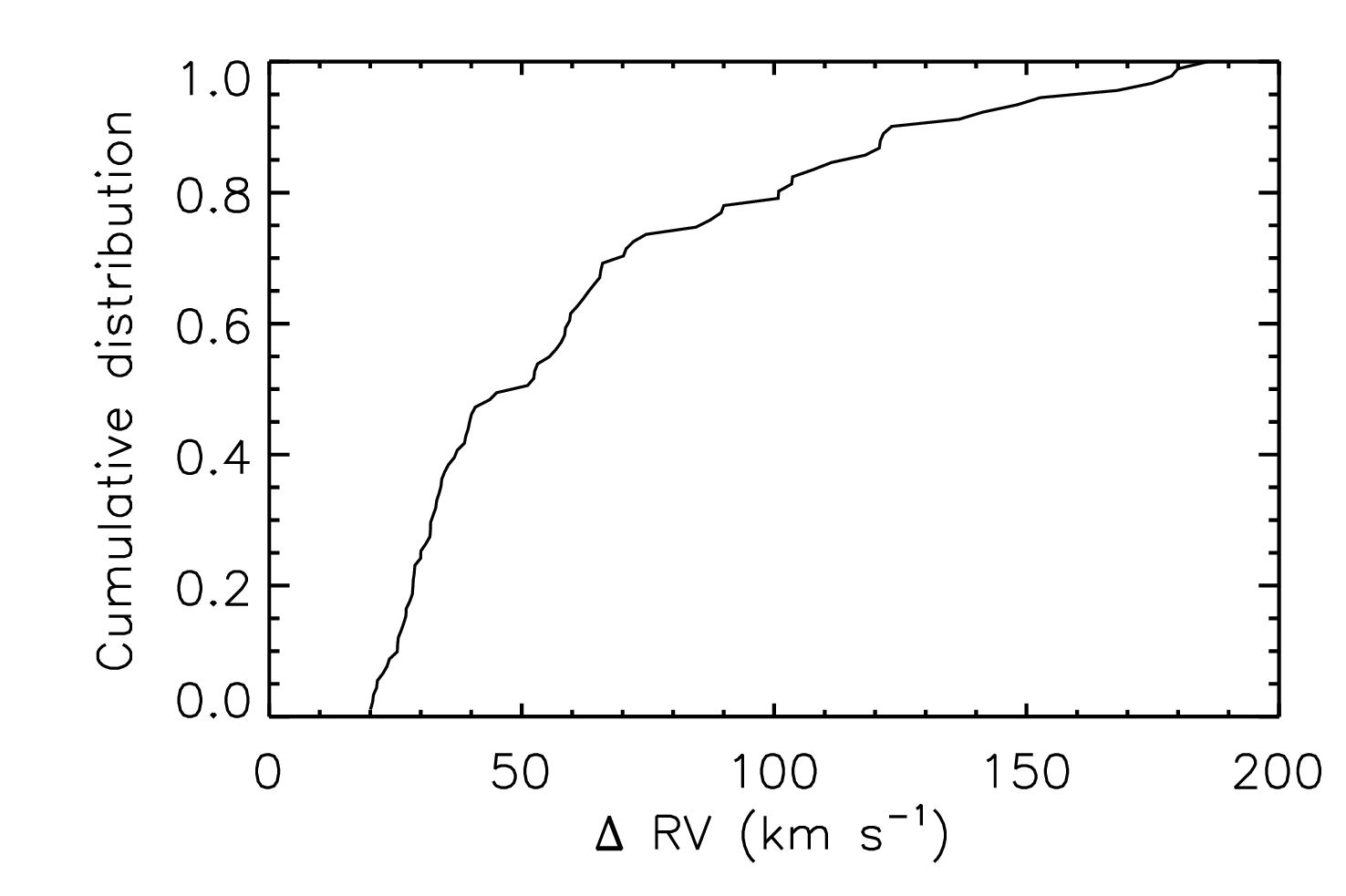}
\includegraphics[width=7.5cm]{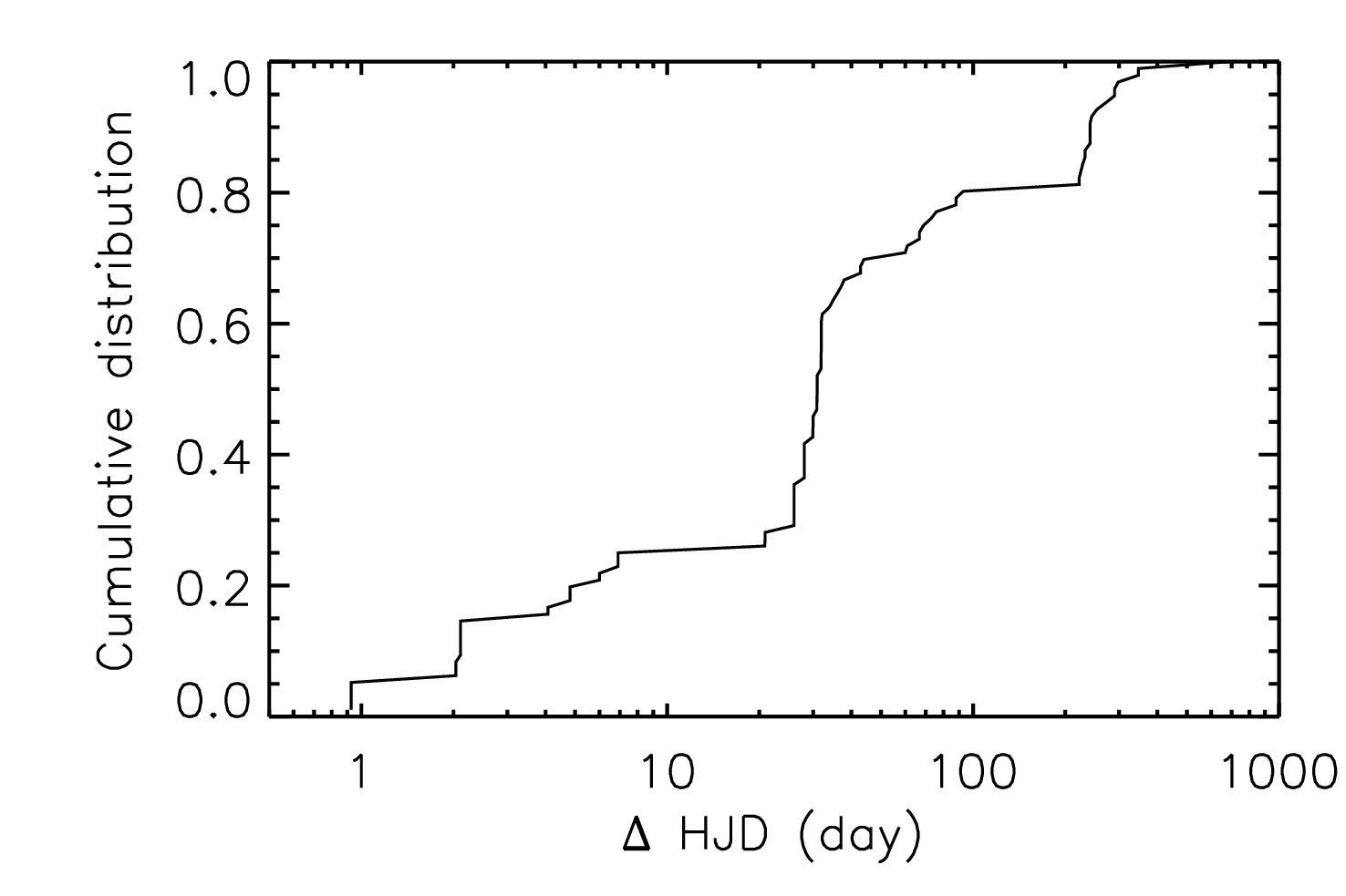}
\caption{Cumulative fractions of the maximum RV amplitude detected
  (upper panel) and the minimum time difference corresponding to
  significant RV-variations (lower panel) for the detected binaries in
  the B-type sample.}\label{f_deltas}
\end{center}
\end{figure}

The upper panel of Figure~\ref{f_deltas} shows the cumulative
distribution of the maximum RV variation for our objects with
estimates in Tables~\ref{RVs} and \ref{RVs_single} which are
classified as binaries.  Approximately $\sim$20\% of the objects in
the binary sample have $\Delta$RV\,$>$\,100\,\kms; these are probably
short-period binaries.

The lower panel of Figure~\ref{f_deltas} shows the minimum time
difference for which significant RV-variations are detected.  The
minimum elapsed time between the LR02 observations of eight of the
nine fibre configurations was typically between one and seven days
(except for Field~C, with a minimum separation of 28 days, see
Appendix of Paper~I); hence we expect that the RV detectability is principally
determined by the properties of the stars rather than the
cadence of the observations. Variations are seen over timescales of one to ten days
for $\sim$30\% of the sample (although this should not be taken as a
direct measure of their orbital periods), while a relatively small
number of the detected systems ($\simeq$15\%) only show variations
over several hundreds of days. Full characterisation of the individual
orbital properties of the detected binaries will require comprehensive
multi-epoch spectroscopy.

\subsection{Spatial variations} \label{sect: spatial_var}
We searched for variations in the observed binary fraction in the
different structures in the region, i.e. NGC\,2070,
NGC\,2060, Hodge\,301, SL\,639 \citep[adopting the same definitions
as those used by][]{eva15}, and the field population. The binary fractions
and numbers of targets in each subsample are summarised in
Table~\ref{tab:f_obs}.  The observed fractions are marginally
larger in NGC\,2070 and NGC\,2060, but remain within the estimated
uncertainties for those in the field population. For the two
older clusters, the observed fractions for the unevolved targets
appear to be slightly lower.

\begin{table*}
\centering
\caption{Observed binary fraction for various subpopulations in the 30~Dor region.}\label{tab:f_obs}
\begin{tabular}{lcccccc}
\hline
Region & \multicolumn{2}{c}{Non-supergiants} & \multicolumn{2}{c}{Supergiants} & \multicolumn{2}{c}{All} \\ 
& $N_\mathrm{obj}$ &$f_{\rm B}$(obs) &$N_\mathrm{obj}$ &$f_{\rm B}$(obs) &$N_\mathrm{obj}$ &$f_{\rm B}$(obs) \\
\hline
NGC\,2070  & \pp93 & 0.27\,$\pm$\,0.05 &   16 & 0.31\,$\pm$\,0.12 &  109 & 0.28\,$\pm$\,0.04 \\
NGC\,2060  & \pp29 & 0.34\,$\pm$\,0.05 & \pp3 & 0.00\,$\pm$\,0.11 &\pp32 & 0.31\,$\pm$\,0.08 \\
Hodge\,301 & \pp12 & 0.08\,$\pm$\,0.08 & \pp3 & 0.00\,$\pm$\,0.27 &\pp15 & 0.07\,$\pm$\,0.06 \\
SL\,639    & \pp10 & 0.10\,$\pm$\,0.09 & \pp4 & 0.25\,$\pm$\,0.27 &\pp14 & 0.14\,$\pm$\,0.09 \\ 
Field      &   217 & 0.24\,$\pm$\,0.03 &   21 & 0.24\,$\pm$\,0.22 &  238 & 0.25\,$\pm$\,0.03 \\
\hline 
All        &   361 & 0.25\,$\pm$\,0.02 &   47 & 0.23\,$\pm$\,0.06 &  408 & 0.25\,$\pm$\, 0.02 \\
\hline
\end{tabular}
\end{table*}

\subsection{Brightness variations} \label{sect: bf_vs_Kmag}
\begin{figure}
\begin{center}
\includegraphics[width=\columnwidth]{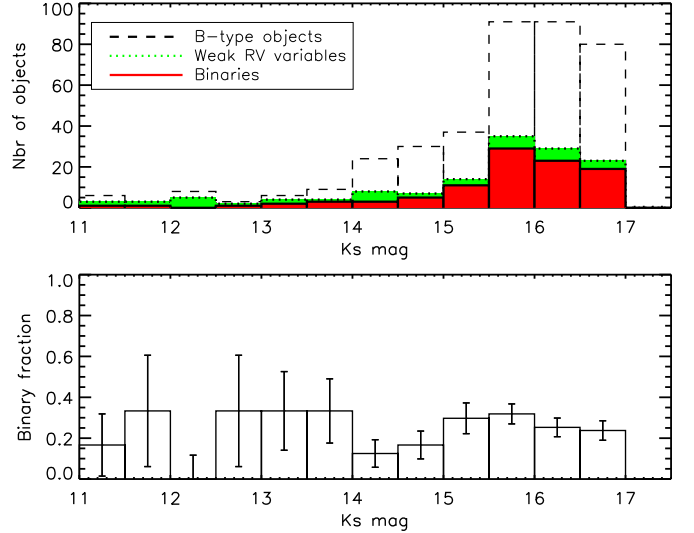}
\caption{Observed binary fraction as a function of $K_{\rm s}$
  magnitude.}\label{f_Ksmag}
\end{center}
\end{figure}

We also investigated the binary fraction as a function of the
brightness of our targets, employing $K_{\rm s}$-band magnitudes from
the VISTA Magellanic Clouds Survey \citep{vmc}. As shown in
Fig.~\ref{f_Ksmag}, there are no obvious trends, suggesting that the
binary fraction is mostly uniform over more than two
orders of magnitude in brightness. 

\section{Intrinsic multiplicity properties} \label{s_model}

The observed binary fraction of the 408 B-type stars is $f_{\rm
  B}$(obs)\,$=$\,0.25 (i.e. 96 positive detections through RV
variations and 5 SB2 system). Because of the cadence of the FLAMES
observations, the S/N of our data, and the orientation of the binary
systems with respect to our line of sight, some binaries will have
eluded detection. Thus, ignoring statistical uncertainties, $f_{\rm
  B}$(obs) represents a lower limit on the true binary fraction of
B-type stars in 30~Dor.

To estimate the intrinsic binary fraction we modelled our criteria for
binarity and estimated our observational biases and detection
sensitivities. In this analysis we ignored the five SB2 systems
discussed in Section~\ref{s_obs} as their double-lined nature
prevented us from obtaining reliable estimates of the RV variation of
the primary stars. Therefore, as a result of this approximation, the
intrinsic binary fraction that we recover may be lower than the true
value by a few percentage points. However, this remains well within
the uncertainty of the method (see below).
 
As discussed in \citetalias{san13}, the detection probability of a
given binary system depends on the properties of the data (uncertainty
of RV estimates, time sampling) and on the properties of the RV signal
that one is trying to detect.  The latter is directly determined by
the orbital properties of the system. The (in)completeness of the VFTS
campaign therefore depends on the distribution of the orbital parameters of
the parent binary population (predominantly the orbital periods, mass
ratios, and eccentricities).  Unfortunately, these quantities are
unknown.  Thus, to constrain the intrinsic multiplicity properties of
the binaries in our sample, we follow the approach implemented in
\citetalias{san13} for the analysis of the O-type stars.

Using a Monte Carlo method, we synthesised populations of stars with
specified parent orbital distributions. Taking into account the
measurement uncertainties and the time sampling of each object, we
sought to reproduce the observational quantities in three aspects:
(i) the observed binary fraction $f_{\rm B}$(obs), (ii) the cumulative
distribution of the maximum amplitude, CDF($\Delta$RV), of significant
RV variations (see Fig.~\ref{f_deltas}, upper panel) and (iii) the
cumulative distribution of the minimum time separation,
CDF($\Delta$HJD), between any pair of RV points that differ
significantly from one another (see Fig.~\ref{f_deltas}, lower panel).

The quality of the match between observations and simulations was
assessed using a global merit function ($\Xi'$).  This was defined as
the product of the Kolmogorov-Smirnov (KS) probabilities between
the synthetic CDFs for $\Delta$RV and $\Delta$HJD and their 
observed distributions, and of the Binomial probability that describes
the chance to obtain the same number of detected binaries
($N_{\rm bin}$) as in our sample, given the {\it simulated} observed binary
fraction $f_{\rm B}({\rm obs})^{\rm simul}$ and the sample size $N$:
\begin{equation}
\Xi' = P_{\rm KS}(\Delta{\rm RV}) \times P_{\rm KS}(\Delta {\rm HJD}) \times
  B(N_{\rm bin},N,f_{\rm B(obs)}^{\rm simul}). \label{eq:merit}
\end{equation}

The method is described further in \citetalias{san13}. Stellar masses
in the simulated populations were randomly drawn from a mass-function
over a given stellar range. We can reasonably assume that the
unevolved stars follow a standard mass-function \citep[e.g.][]{kro01},
but the situation may be different for the supergiants, many of which
would have been born as higher-mass stars, which could have
experienced considerable evolution prior to their current phase (e.g.
mass lost via stellar winds, binary interactions). Therefore, we focus
on the 357 unevolved B-type stars in our sample (which does not
include the four SB2 systems with unevolved stars; cf. the numbers in
Table~\ref{tab:f_obs}).

As in \citetalias{san13}, we considered orbital periods ($P$),
mass-ratios ($q=M_2/M_1$) and eccentricities ($e$) over ranges of
0.15\,$\leq$\,$\log (P/{\rm d})$\,$\leq$\,3.5,
0.1\,$\leq$\,q\,$\leq$\,1.0 and 0\,$<$\,e\,$\leq$\,0.9, respectively.
We adopted power-laws to describe the distributions of the orbital
parameters: $f_{\rm P}$\,$\propto$\,$(\log P/{\rm d})^\pi$, $f_{\rm
  q}$\,$\propto$\,$q^\kappa$ and $f_{\rm e}$\,$\propto$\,$e^\eta$. We
varied $\pi$, $\kappa$ and the intrinsic number of binaries to best
reproduce the observed quantities CDF($\Delta$RV), CDF($\Delta$HJD)
and $f_{\rm B}$(obs). Our method was insensitive to the eccentricity
distribution and we could not constrain $\eta$. As in our analysis of
the O-stars, we adopted the value obtained from O-type binaries in
Galactic open clusters, $\eta$\,$=$\,$-$0.5 \citep{san12}. While there
is no guarantee that this value is appropriate, it preserves the
homogeneity of the analysis of the O- and B-type populations from the
VFTS.  The range of values explored for $f_{\rm B(true)}$, $\pi$ and
$\kappa$ are given in Table~\ref{tab:MCgrid}, and the criteria to
identify binarity in the Monte Carlo simulations are those described
in Section~\ref{s_bc}.

\begin{table*}
\centering
\caption{Properties of the Monte Carlo grid. The first three 
  rows provide information on the orbital distributions: physical parameter 
  (Col.~1), probability-density function (Col.~2) and applicability domain 
  (Col.~3). Cols.~4 to 6 give the quantity, investigated range and step 
  size in the grid.}
\label{tab:MCgrid}
\begin{tabular}{lllccc}
\hline 
\hline 
Parameter &  pdf               & Domain    & Var.          &   Range           & Step \\
\hline 
$P$ (d) & $(\log_{10} P)^{\pi}$ & 0.15 - 3.5    & $\pi$     & $-$2.50 - $+$2.50  & 0.1 \\
$q$     & $q^{\kappa}$          &  0.1 - 1.0    & $\kappa$  & $-$3.50 - $+$1.50 & 0.1 \\  
$e$     & $e^{\eta}$            &  $10^{-5}$ - 0.9 & $\eta$    & $-$0.5 (fixed)    & $n/a$     \\   
$f_{\rm B(true)}$ & $n/a$       & $n/a$         & $f_{\rm B(true)}$ & $+$0.40 - $+$1.00 & 0.02 \\  
\hline 
\end{tabular}  
\end{table*}

\begin{figure*}
\centering
\includegraphics[width=\textwidth]{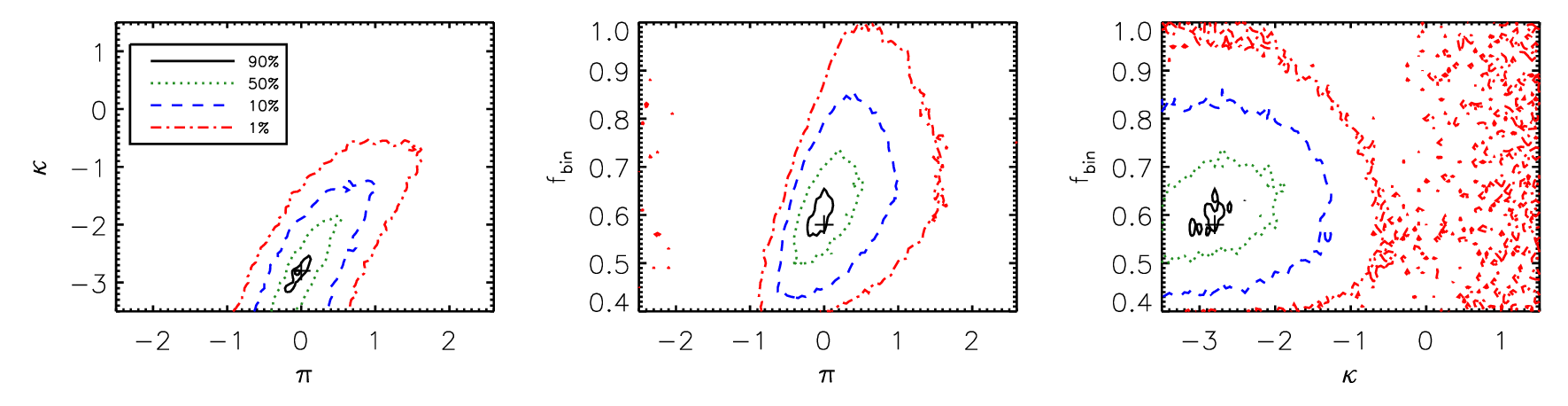}
\caption{Projections of the global merit function, $\Xi'$, on the
  pairs of planes defined by $\pi$, $\kappa$ and $f_{\rm B(true)}$.  The location
  of the absolute maximum is indicated by a cross ($+$) and contours
  indicate loci of equal values in the merit function, expressed as a
  fraction of its absolute maximum (see inset in left-hand
  panel).}\label{fig:merit_2d}

 \includegraphics[width=\textwidth]{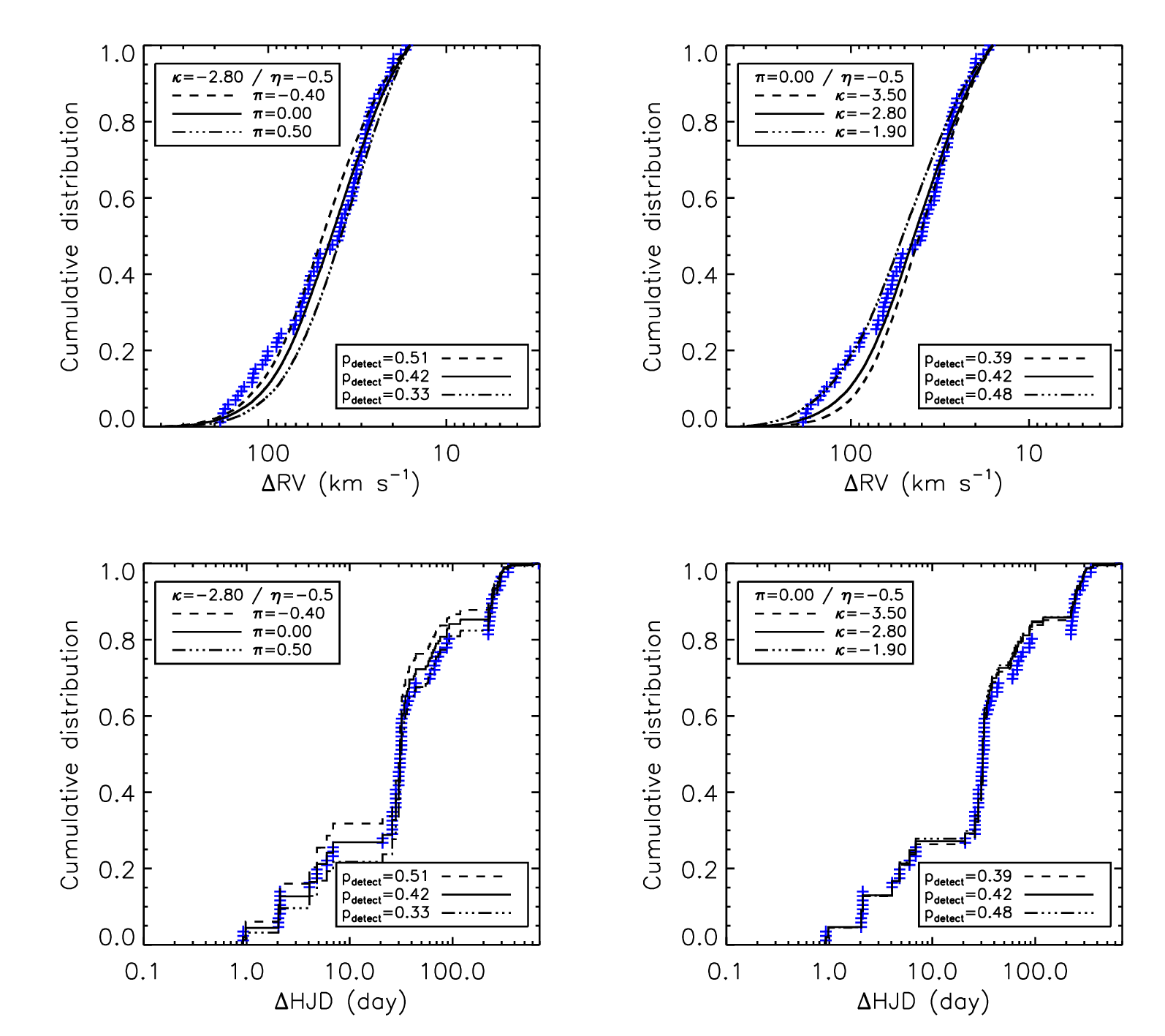}
 \caption{Comparison of the observed (crosses) and simulated (lines)
   cumulative distributions of the peak-to-peak RV amplitudes (upper
   panels) and of the variability timescales (lower panels).  The
   exponent $\pi$ (period distribution) is varied in the left-hand
   panels, while $\kappa$ (the exponent of the mass-ratio distribution),
   is varied in the right-hand panels. Values of $\pi$, $\kappa$\ and
   $\eta$ are indicated in the upper-left insets of each panel. The
   lower-right insets indicate the VFTS detection probability
   for the range of parameters specified in Table~\ref{tab:MCgrid} and
   for the adopted parent distributions.}  \label{fig:bestfit}
\end{figure*}

Fig.~\ref{fig:merit_2d} shows the behaviour of $\Xi'$ when varying
$f_{\rm B(true)}$, $\pi$ and $\kappa$ and Fig.~\ref{fig:bestfit} shows the
simulated distributions for the optimum $\Xi'$ compared with the
observations. The best representation of the data is obtained with
$f_{\rm B(true)}$\,$=$\,0.58\,$\pm$\,0.11, $\pi$\,$=$\,0.0\,$\pm$\,0.5 and
$\kappa$\,$=$\,$-2.8$\,$\pm$\,0.8. 

We estimated the uncertainties on the retrieved multiplicity
properties by the use of 50 synthetic datasets which share the same
properties as our observations in terms of RV measurement
uncertainties and time sampling, and that are drawn from known
intrinsic distributions (see Appendix~\ref{app:MCerror}).  These tests
also enabled us to look for systematics in our methods. From
Table~\ref{tab:mctest}, our approach tends to underestimate the
intrinsic binary fraction by a couple of percentage points.  The value
obtained for the index of the period distribution $\pi$ is typically
0.2~dex smaller than the adopted index. Similarly, the retrieved index
of the mass ratio distribution $\kappa$ is typically 0.5~dex smaller
than the adopted value. Nonetheless, these systematics remain smaller
than the estimated statistical uncertainties on the measured value.

We note that the $\kappa$ value estimated here differs to those found
for Galactic binaries, e.g.  $\kappa$\,$=$\,$-$0.1\,$\pm$\,0.6
\citep{san12} and $\kappa$\,$\sim$\,0 \citep{k14}, and
$\kappa$\,$=$\,$-$1.0\,$\pm$\,0.4 for the O-type stars from the VFTS
\citepalias[]{san13}; this remains the case when taking into account
the apparent bias to smaller values from our method
(Appendix~\ref{app:MCerror}), as well as the omission of the five SB2
systems that will create an additional bias in the same sense.
However, as discussed in Appendix~C of \citetalias{san13}, the
Monte Carlo method is only weakly sensitive to the adopted value of
$\kappa$, with large uncertainties on its estimated value.
Higher-quality spectroscopy (both in terms of S/N and cadence) of the
B-type stars studied in this work would enable a more rigorous
analysis of the mass-ratio distribution in the detected binaries.

From these calculations, we estimate that the overall detection
probability of B-type binaries from the VFTS campaign is
40\,$\pm$\,10\%. This suggests that we detected fewer than half the
binaries with a B-type primary component.  For the detected 
binary fraction $f_{\rm B(obs)}$ and a detection probability $p$, a fraction
[$f_{\rm obs}\,\times\,(1-p)$]\,/\,[$p(1-f_{\rm obs})$] $\simeq$
45\% of these `single' stars are expected to be binary.

\section{Summary}\label{s_conclusions}

We have investigated the multiplicity properties of 408 B-type stars
in 30~Dor, employing the multi-epoch spectroscopy from the VFTS.  We
find that at least 25\% of the stars display significant
RV variations that are large enough to potentially be the
spectroscopic signature of binarity. An additional 9\% of the stars
display RV variations -- these are either binary
candidates or display intrinsic atmospheric variability.

By modelling the properties of our observations, we estimate our
observational biases and constrain the intrinsic multiplicity
properties of the B-type stars in 30~Dor. The binary fraction of the
unevolved (i.e. dwarf/giant) stars is estimated to be $f_{\rm
  B(true)}$\,$=$\,0.58\,$\pm$\,0.11, with a flat period distribution.
The uncertainties on the mass-ratio distribution are large and the
best-fit distribution should be considered with caution.  Nonetheless,
the multiplicity properties of the B-type stars agree (within the
uncertainties) with those of the O-type stars in 30~Dor.  Although the
B-star sample is larger, the uncertainties are greater than for the O
stars, a consequence of the B stars being fainter (hence having
spectroscopy with generally lower S/N).  Our results indicate that the
important evolutionary implications of binarity inferred for the
O-star sample probably also hold for the stars born as B-type, and
hence for the entire population of supernova progenitors.
Comprehensive spectroscopic monitoring of the B-type sample is now
required to determine the period distribution to firmly test this
conclusion.

\begin{acknowledgements}
  Based on observations at the European Southern Observatory Very
  Large Telescope in programme 182.D-0222; we are grateful to the ESO
  staff at Paranal with their assistance in obtaining the data.  We
  thank the referee, Dr Andrei Tokovinin, for his constructive
  comments on the manuscript. SdM acknowledges support by a Marie
  Sk{\l}odowska-Curie Reintegration Fellowship (H2020-MSCA-IF-2014,
  project id. 661502) awarded by the European Commission.
  Additionally, we acknowledge financial support from the UK Science
  and Technology Facilities Council, the Netherlands Science
  Foundation, the Leverhulme Trust and the Department of Education and
  Learning in Northern Ireland.
\end{acknowledgements}

\bibliography{26192}

\Online
\begin{appendix}

\section{Spatially-resolved companions}\label{visual_checks}

The Medusa fibres subtend 1\farcs2 on the sky which, at the distance
of the LMC (50\,kpc), is equivalent to 0.29\,pc. Some of the spectra
could therefore be composites of both wide binaries and/or chance
line-of-sight alignments. High-quality optical imaging from the
wide-field F775W mosaic of 30~Dor taken with the {\em Hubble Space
  Telescope (HST)} in programme GO-12499 \citep[PI: Lennon;
see][]{sab13} includes 300 of the 403 stars (74\%) with RV estimates.

We checked the available shallow and deep images for companions within
1$''$.  Two of us (PLD and SdM) visually checked each of our targets for
nearby companions which could have significantly contaminated the LR02
spectroscopy; notable companions/features in this context are
summarised in Table~\ref{visual}. Although subjective, the two
independent evaluations were in excellent agreement, ensuring that
important companions, as resolved by {\em HST} imaging, are noted.

We detect visual companions that could contribute to the observed
spectra for thirty-five of the B-star targets (i.e. 11.7\,\% of those
with imaging available).  In the context of searching for
spectroscopic binaries such visual companions sample a very different
range of physical separations \citep[see e.g. Fig.~1 from][]{se11}
and they will not influence the results for the {\em spectroscopic}
binary sample analysed in this paper. Even if they are physically
bound, the RV amplitudes for these visual-binary systems will be below
our detection threshold, and they will therefore be categorised as
single using the spectroscopic criteria described in
Section~\ref{s_analysis} (in the absence of other sources of RV
variability).  Such wide systems are unlikely to interact at any point
in their evolution, and so will have no influence on our conclusions
in this respect.  However, to the extent that the observed spectra
will be composite to some (generally small) degree, this contamination
will need to be taken into account in any future atmospheric analyses
of these spectra.

\begin{table}[h]
  \caption{B-type targets that may have a significant 
    contribution to their FLAMES spectra from nearby companions.}\label{visual}
\begin{tabular}{cl}
\hline
VFTS & Comment \\
\hline
043 &  Elongated image - close companion  \\
044 &  Elongated image - close companion  \\
050 &  Star with similar mag. at 1\arcsec  \\
068 &  Fainter star at $\sim$0.5\arcsec \\
127 &  Fainter star at $<$0.5\arcsec \\
133 &  Star with similar mag. at $\sim$0.5\arcsec \\
167 &  Elongated image - close companion  \\
212 &  Elongated image - close companion  \\
238 &  Two fainter stars, at $<$0.5\arcsec \& $\sim$0.5\arcsec \\
276 &  Star with similar mag. at $\sim$0.5\arcsec \\
278 &  Crowded field, 4 fainter stars at $<$0.5\arcsec \\
283 &  Two stars with similar mag. at $\sim$0.5\arcsec \\
292 &  Three fainter stars at $<$0.5\arcsec \\
301 &  Star with similar mag. at $\sim$1\arcsec \\
374 &  Fainter star at $<$0.5\arcsec \\
376 &  Elongated image - close companion \\
381 &  Star with similar mag. at $\sim$0.5\arcsec \\
403 &  Star with similar mag. at $\sim$1\arcsec \\
447 &  Star with similar mag. at $<$0.5\arcsec \\
448 &  Star with similar mag. at $<$0.5\arcsec \\
461 &  Star with similar mag. at $<$0.5\arcsec \\
463 &  Fainter star at $<$0.5\arcsec \\
480 &  Two fainter stars at $<$0.5\arcsec \\
504 &  Star with similar mag. at $\sim$1\arcsec \\
548 &  Fainter star at $<$0.5\arcsec \\
553 &  Fainter star at $\sim$0.5\arcsec \\
567 &  Crowded field, star with similar mag. at 0.5\arcsec \\
584 &  Two fainter stars at $\sim$0.5\arcsec \\
632 &  Fainter star at 0.5\arcsec \\
643 &  Star with similar mag. at $\sim$0.5\arcsec \\
644 &  Star with similar mag. at $\sim$0.5\arcsec \\
662 &  Fainter star at $<$0.5\arcsec \\
712 &  Star with similar mag. at $<$0.5\arcsec \\
747 &  Star with similar mag. at $<$0.5\arcsec \\
833 &  Elongated image - close companion  \\
\hline
\end{tabular}
\end{table}

\section{Accuracy of the MC method}\label{app:MCerror}

As in \citetalias{san13}, we estimated the accuracy of the method by
applying it to synthetic data which shared the same properties as the
sample of early-B dwarfs analysed in this paper. We adopted different
parent distributions to generate the synthetic observations and
investigated how well the method behaved in different parts of
parameter space, including the region where the adopted parameters led
to the best representation of the merit function.  The results from
these tests are reported in Table~\ref{tab:mctest}.

\begin{table*}
\centering
\caption{Overview of test results from synthetic datasets. The input multiplicity 
  parameters are given first in Cols.~2 to 4. The medians and 0.16 and 0.74 percentiles 
  of the retrieved parameters from a set of 50 test runs are then indicated. 
  The properties of the computed Monte Carlo grid are the same as those in 
  Table~\ref{tab:MCgrid}.}\label{tab:mctest}
\begin{tabular}{llcc}
\hline \\
$\Delta$RV$_\mathrm{min}$ & \multicolumn{3}{c}{Multiplicity properties} \\
{[\kms]}                  & $f_{\rm B(true)}$ & $\pi$ & $\kappa$                    \\
\hline \\
      & ($f_{\rm B(true)}$\,$=$\,0.70) & ($\pi_\mathrm{true}=0.0$) & ($\kappa_\mathrm{true}=0.0$)  \\
16.0  &   0.70  [ 0.62,0.82] & $-$0.20  [$-$0.50,$+$0.20] & $-$0.40  [$-$1.00,$+$0.90]  \\ \\         
      & ($f_{\rm B(true)}$\,$=$\,0.70) & ($\pi_\mathrm{true}=-0.5$) & ($\kappa_\mathrm{true}=0.0$)  \\     
16.0  &   0.68  [ 0.62,0.76] & $-$0.70  [$-$1.10,$-$0.40] & $-$0.50  [$-$1.00,$+$0.60]  \\ \\         
      & ($f_{\rm B(true)}$\,$=$\,0.70) & ($\pi_\mathrm{true}=0.0$) & ($\kappa_\mathrm{true}=-2.5$)  \\    
16.0  &   0.68  [ 0.54,0.78] & $-$0.20  [$-$0.50,$+$0.20] & $-$2.90  [$-$3.50,$-$2.20]  \\\\         
      & ($f_{\rm B(true)}$\,$=$\,0.55) & ($\pi_\mathrm{true}=0.0$) & ($\kappa_\mathrm{true}=-2.5$)  \\    
16.0  &   0.52  [ 0.42,0.64] & $-$0.20  [$-$0.60,$+$0.40] & $-$3.00  [$-$3.50,$-$2.00] \\
\hline 
\end{tabular}
\end{table*}

\end{appendix}

\end{document}